\documentclass[iop, twocolumn]{aastex63}
\usepackage{verbatim,subfigure,amsmath}
\hypersetup{citecolor=blue,urlcolor=red}
\shorttitle{Radio pulse search}
\shortauthors{Peng et al.}

\begin{document}

\title{Radio pulse search from Aql X-1}

\correspondingauthor{S.Q. Wang, Zhaosheng Li}
\email{wangshuangqiang@xao.ac.cn, lizhaosheng@xtu.edu.cn}

\author{Long Peng}
\affiliation{Key Laboratory of Stars and Interstellar Medium, Xiangtan University, Xiangtan 411105, Hunan, People’s Republic of China}
\affiliation{Xinjiang Astronomical Observatory, Chinese Academy of Sciences, Urumqi 830011, Xinjiang, People’s Republic of China}

\author[0000-0003-2310-8105]{Zhaosheng Li}
\affiliation{Key Laboratory of Stars and Interstellar Medium, Xiangtan University, Xiangtan 411105, Hunan, People’s Republic of China}

\author{Yuanyue Pan}
\affiliation{Key Laboratory of Stars and Interstellar Medium, Xiangtan University, Xiangtan 411105, Hunan, People’s Republic of China}

\author[0000-0001-7595-1458]{Shanshan Weng}
\affiliation{Department of Physics and Institute of Theoretical Physics, Nanjing Normal University, Nanjing 210023, Jiangsu, People’s Republic of China}

\author[0000-0002-7662-3875]{Wenming Yan}
\affiliation{Xinjiang Astronomical Observatory, Chinese Academy of Sciences, Urumqi 830011, Xinjiang, People’s Republic of China}
\affiliation{Key Laboratory of Radio Astronomy, Chinese Academy of Sciences, Urumqi 830011, Xinjiang, People’s Republic of China}
\affiliation{Xinjiang Key Laboratory of Radio Astrophysics, Urumqi 830011, Xinjiang, People’s Republic of China}

\author[0000-0002-9786-8548]{Na Wang}
\affiliation{Xinjiang Astronomical Observatory, Chinese Academy of Sciences, Urumqi 830011, Xinjiang, People’s Republic of China}
\affiliation{Key Laboratory of Radio Astronomy, Chinese Academy of Sciences, Urumqi 830011, Xinjiang, People’s Republic of China}
\affiliation{Xinjiang Key Laboratory of Radio Astrophysics, Urumqi 830011, Xinjiang, People’s Republic of China}

\author[0000-0002-9434-4773]{Bo-Jun Wang}
\affiliation{National Astronomical Observatories, Chinese Academy of Sciences, Beijing 100101, People’s Republic of China}

\author[0000-0003-4498-6070]{Shuangqiang Wang}
\affiliation{Xinjiang Astronomical Observatory, Chinese Academy of Sciences, Urumqi 830011, Xinjiang, People’s Republic of China}
\affiliation{CSIRO Astronomy and Space Science, P.O. Box 76, Epping, NSW 1710, Australia}
\affiliation{Key Laboratory of Radio Astronomy, Chinese Academy of Sciences, Urumqi 830011, Xinjiang, People’s Republic of China}
\affiliation{Xinjiang Key Laboratory of Radio Astrophysics, Urumqi 830011, Xinjiang, People’s Republic of China}

\begin{abstract}

We present 12 observations of the accreting millisecond X-ray pulsar Aql X-1, taken from 2022 August to 2023 October using the Five-hundred-meter Aperture Spherical Radio Telescope at 1250 MHz. 
These observations covered both the quiescence and X-ray outburst states, as determined by analyzing the X-ray data from the Neutron Star Interior Composition Explorer and the Monitor of All-sky X-ray Image. 
Periodicity and single-pulse searches were conducted for each observation, but no pulsed signals were detected. The obtained upper limit flux densities are in the range of $2.86-5.73~{\rm \mu Jy}$, which provide the lowest limits to date. 
We discuss several mechanisms that may prevent detection, suggesting that Aql X-1 may be in the radio-ejection state during quiescence, where the radio pulsed emissions are absorbed by the matter surrounding the system.

\end{abstract}


\section{Introduction}

Millisecond pulsars (MSPs) are a subclass of pulsars characterized by short rotation periods (e.g., $< 30\,$ms) and low magnetic field strengths ($\sim 10^8 \,$G; see \citealt{Bhattacharya1991, Lorimer2004, Manchester2004,Manchester2017,Lorimer2005} for a review). 
The first MSP, PSR B1937+21, was discovered in 1982 by the Arecibo radio telescope~\citep{Backer1982}. 
Shortly after the discovery of MSPs, the recycling scenario was proposed \citep{Alpar1982,Radhakrishnan1982}. In this model, pulsars are spun up to millisecond periods by accreting matter and angular momentum from a companion star in low-mass X-ray binaries (LMXBs). 
During the accretion process, the neutron star is detectable as an X-ray source, known as an accreting millisecond X-ray pulsar (AMXP). 
When the companion star decouples from its Roche lobe, accretion in the AMXP ceases, as does the X-ray emission generated by mass transfer. At this stage, the radio emission from the MSP becomes active. 
According to the recycling scenario, some LMXBs should host neutron stars with millisecond rotation periods.  
Until 1998, \citet{Wijnands1998} provided the first confirmation of the recycling scenario by detecting coherent pulsations with a period of 2.5\,ms in the LMXB SAX J1808.4$-$3658. 
The recycling scenario has been further supported by the discovery of transitional MSPs (tMSPs), which transition between rotation-powered and accretion-powered states, thereby bridging the gap between radio MSPs and AMXPs~\citep{Archibald2009,Papitto2013,Bassa2014}. 

More than 20 AMXPs have been discovered since the first detection of SAX J1808.4$-$3658~\citep{Wijnands1998,Di2020,Patruno2021}. 
AMXPs are typically X-ray transients, with X-ray luminosities of $\sim10^{36}-10^{37} \, \text{erg} \, \text{s}^{-1}$ during outbursts, separated by extended quiescent periods at $L_X \sim 10^{31} - 10^{32} \, \text{erg} \, \text{s}^{-1}$ (e.g., see \citealt{Patruno2021} for a review). 
According to the recycling scenario, AMXPs are expected to behave as rotation-powered radio MSPs during X-ray quiescence and as accretion-powered X-ray MSPs during outburst states~\citep{Campana1998}. 
However, no radio pulsed emissions have been detected during the quiescence of AMXPs, including sources such as Aql X-1~\citep{Burgay2003}, XTE J0929$-$314~\citep{Iacolina2009}, XTE J1751$-$305, XTE J1814$-$338, and SAX J1808.4$-$3658~\citep{Iacolina2010}.
\citet{Burderi2001} proposed the radio-ejection model, in which an active radio pulsar during quiescence ejects accreting material from the system due to the pressure of its emission. However, the pulsar's radio emissions are undetectable because of absorption by surrounding material, and searching for radio pulses at higher frequencies is suggested. 
Alternatively, for systems with long orbital periods, the
ejected matter is spread over a wide orbit, which may reduce the amount of free-free absorption, increasing the likelihood of radio detection~\citep{Di2020}. 

Aql X-1 was discovered in 1965~\citep{Friedman1967}, with an orbital modulation of 18.95 hr and an orbital inclination of $36^\circ\text{--}47^\circ$~\citep{Chevalier1991, Welsh2000}. Its distance is estimated to be between 4.0 and 5.75 \text{kpc}~\citep{Li2017}. The optical counterpart of Aql X-1, designated as V1333 Aql, was identified in 1978~\citep{Thorstensen1978}.  
Aql X-1 is an intermittent AMXP, with coherent pulsations at a frequency of 550.273(1) Hz detected for only 120 s during a total exposure time of 1645 ks~\citep{Casella2008}. 
Since its discovery, Aql X-1 has undergone multiple outbursts, approximately once a year, typically lasting from weeks to months~\citep{Degenaar2019}.  

The radio counterpart of Aql X-1 has been detected during some outbursts (e.g., \citealt{Tudose2009,Miller2010,2018A&A...616A..23D,Motta2019}). 
Searches for radio pulsed emissions have been conducted using the 76-m Lovell Telescope at 925 MHz~\citep{Biggs1996} and the Parkes 64-m radio telescope at 1.4\,GHz~\citep{Burgay2003}, but no pulsed signals were detected, with upper flux density limits of a few mJy.  

Due to its high sensitivity, the Five-hundred-meter Aperture Spherical radio Telescope (FAST; \citealt{Nan2011}) is an ideal instrument for searching for radio pulsed emissions from AMXPs.  
Here, we report on a search for radio pulsed emissions from Aql X-1 using FAST. 
We also note that Aql X-1 has a longer orbital period of 19 hr compared to other AMXPs~\citep{Di2020}, which may increase the likelihood of detection according to the radio-ejection model~\citep{Burderi2001}.  
In Section \ref{sec:obs}, we introduce the observations and data processing methods. 
In Section \ref{sec:results}, we present the results of the radio pulse search.  
In Section \ref{sec:con}, we discuss and summarize our findings.

\section{Observations and Data Processing}
\label{sec:obs}

\subsection{FAST Observations}

We conducted 12 observations of Aql X-1 with FAST -- 2 in 2022 and 10 in 2023 (Table~\ref{tab:observation}). These observations were used the central beam of the 19-beam receiver, covering a frequency range of 1.05--1.45\,GHz, with a central observing frequency of 1.25\,GHz~\citep{Jiang2020}. The data were recorded in the PSRFITS format~\citep{Hotan2004}, with 8-bit resolution, 4096 frequency channels, each having a bandwidth of 0.122\,MHz, four polarizations, and a sampling interval of 49.152\,$\mu$s.

\begin{table*}
\centering
\caption{Details of FAST Observation and Data Reduction}
\label{tab:observation}
\renewcommand{\arraystretch}{1}  
\setlength{\tabcolsep}{8pt}
\begin{tabular}{ccccccccc}
\toprule
Obs.No.  & Obs. Date &  Obs. MJD   & Length   &Orbital Phase  &  \(z_{\text{max}}\) &  \(z_{\text{max}}^\mathrm{(a)}\) & State  & Sensitivity    \\
&  &  &  (hr)    &  &  & &    & $(\mu\text{Jy})$   \\
\hline
01 & 2022-08-31 & 59822.50 & 2.0 & 0.51 - 0.60 & 430 & 600 & q & 2.86 \\
02 & 2022-09-29 & 59851.46 & 1.0 & 0.18 - 0.23 & 108 & 200 & q & 4.05 \\
03 & 2023-06-20 & 60114.76 & 1.0 & 0.68 - 0.79 & 108 & 200 & q & 4.05 \\
04 & 2023-07-02 & 60126.73 & 1.0 & 0.85 - 0.90 & 108 & 200 & q & 4.05 \\
05 & 2023-07-18 & 60143.63 & 1.3 & 0.26 - 0.33 & 191 & 200 & q & 3.51 \\
06 & 2023-08-26 & 60182.59 & 0.5 & 0.60 - 0.63 & 27  & 100 & o & 5.73 \\
07 & 2023-09-02 & 60189.51 & 0.5 & 0.37 - 0.40 & 27  & 100 & o & 5.73 \\
08 & 2023-09-09 & 60196.59 & 0.5 & 0.33 - 0.36 & 27  & 100 & o & 5.73 \\
09 & 2023-09-16 & 60203.50 & 0.5 & 0.09 - 0.11 & 27  & 100 & o & 5.73 \\
10 & 2023-09-24 & 60211.44 & 0.5 & 0.14 - 0.17 & 27  & 100 & q & 5.73 \\
11 & 2023-10-01 & 60218.42 & 0.5 & 0.98 - 0.01 & 27  & 100 & q & 5.73 \\
12 & 2023-10-21 & 60238.38 & 0.5 & 0.27 - 0.30 & 27  & 100 & q & 5.73 \\
\hline
\end{tabular}
\footnotesize{Column (1) observation number, (2) observation date, (3) observation Modified Julian Date (MJD), (4) observation duration, (5) orbital phase, (6) calculated $z_{\text{max}}$ value, (7) actual $z_{\text{max}}$ value used in accelerated search, (8) the outburst (o) or quiescent (q) state, (9) upper limit flux density.} 
\end{table*}

\subsection{Periodicity Search}

We carried out blind periodic searches for the observations of Aql X-1 with FAST using the {\sc PRESTO} software package\footnote{\url{https://github.com/scottransom/presto}}~\citep{Ransom2001}. 
For each observation, we used \texttt{rfifind} to identify radio frequency interference (RFI) in both the time and frequency domains and to generate an RFI mask. 
We then determined the range of trial dispersion measures (DMs) for searching Aql X-1. The observed width of a radio pulse is influenced by propagation effects and signal-processing response times~\citep{Cordes2003}, and can be expressed as
\begin{equation}
\label{eq:width}
W_{\rm obs} = \sqrt {W^2_{\rm int} + t^2 + t^2_{\rm DM} + t^2_{\rm scatt}},
\end{equation}
where \( W_{\rm int} \) is the intrinsic pulse width, \( t \) is the time resolution of the receiver, \( t_{\rm DM} \) is the pulse broadening due to smearing across individual frequency channels, and \( t_{\rm scatt} \) is the broadening due to interstellar scattering. 
For our observations with FAST, \( t \) is small and can be neglected. Both \( t_{\rm DM} \) and \( t_{\rm scatt} \) depend on the DM. The \( t_{\rm DM} \) is given by
\begin{equation}
t_{{\rm DM}} = 8.3 \, {\mu s} \, {\rm DM} \, \Delta \nu_{\rm MHz} \, \nu_{\rm GHz}^{-3},
\label{eq:t_dm}
\end{equation}
where \( \nu_{\rm MHz} \) and \( \Delta \nu_{\rm MHz} \) are the central observing frequency and bandwidth of an individual frequency channel, respectively. The \( t_{\rm scatt} \) can be expressed as~\citep{Cordes2002}
\begin{equation}
\begin{aligned}
\label{eq:scattering}
{\rm log} \, t_{\rm scatt} = -3.72 + 0.411\,{\rm log}\,{\rm DM} + 0.937({\rm log} \, {\rm DM})^2 \\ - 4.4\,{\rm log}\,\nu_{\rm GHz} \, \mu s.
\end{aligned}
\end{equation}

The rotation period of Aql X-1 has been determined in the X-ray wave band as 1.8\,ms~\citep{Casella2008}.
For our observations, \( \nu = 1.25 \,{\rm GHz} \) and \( \Delta \nu = 0.122 \,{\rm MHz} \). 
For a DM of 500\,pc\,cm\(^{-3}\), we found \( t_{\rm DM} = 0.26 \,{\rm ms} \), which is negligible, and \( t_{\rm scatt} = 6.14 \,{\rm ms} \), which is several times of the rotation period of Aql X-1. 
We arbitrarily set the maximum trial DM in our search to 500\,pc\,cm\(^{-3}\). 
Note that the estimated DM of Aql X-1 is 147\,pc\,cm\(^{-3}\), based on the YMW16 electron density model~\citep{Yao2017}.

We used the procedure \texttt{DDplan.py} to determine near-optimal dedispersion methods, setting DM steps of 0.05 and 0.1\,pc\,cm\(^{-3}\) for DMs from 0--370 and 370--500\,pc\,cm\(^{-3}\), respectively. 
The \texttt{prepsubband} tool was used to dedisperse subbands, and \texttt{realfft} performed a discrete Fourier transform (DFT) on the dedispersed time series. 
Since Aql X-1 is in a binary system, we used \texttt{accelsearch} to conduct Fourier-domain acceleration searches. 

The \( z_{\text{max}} \) parameter, which defines the maximum number of Fourier bins that the highest harmonic can drift linearly in the power spectrum~\citep{Ransom2001}, is calculated as
\begin{equation}
z_{\rm max} = \frac{a_{\rm max} T^2 f_0}{c},
\label{eq:zmax}
\end{equation}
where \( T \) is the total integration time, \( f_0 \) is the pulsar's spin frequency, \( a_{\rm max} \) is the maximum orbital acceleration, and \( c \) is the speed of light. 
The calculated \( z_{\rm max} \) for each observation is shown in the sixth column of Table~\ref{tab:observation}. 
Note that the actual \( z_{\rm max} \) may be larger, as Eq.~\ref{eq:maxA} does not account for the eccentricity of the binary orbit. 
The values of \( z_{\rm max} \) used in our search are provided in the seventh column of Table~\ref{tab:observation}.

The \( a_{\text{max}} \) is derived using Kepler's third law:
\begin{equation}
a_{\text{max}} = \left(\frac{2\pi}{P_{\text{orb}}}\right)^{4/3} (T_{\sun} f)^{1/3} c,
\label{eq:maxA}
\end{equation}
where \( P_{\text{orb}} \) is the orbital period, \( T_{\sun} = GM_{\sun}/c^{3} = 4.925490947 \, \mu s \), and \( f \) is the mass function, defined as
\begin{equation}
f = \frac{4\pi^2}{G} \frac{x^3}{P_\mathrm{orb}^2} = \frac{(m_{\rm c} \sin{i})^{3}}{(m_{\rm p} + m_{\rm c})^{2}}.
\end{equation}
Using X-ray reflection modeling, the orbital inclination \( i \) is constrained to \( (26 \pm 2)^\circ \)~\citep{Ludlam2017}, and assuming the pulsar mass \( m_{\text{p}} \) to be 1.4\,\( M_{\sun} \)~\citep{Thorsett1999}, we calculate \( a_{\text{max}} = 4.52 \, \text{m s}^{-2} \).

We used \texttt{ACCEL\_sift.py} to reject bad candidates and combine detections of the same candidate. We also employed \texttt{Jinglepulsar}\footnote{\url{https://www.github.com/jinglepulsar}} for more sensitive detection of faint signals~\citep{jinglepulsar2021}. 
For each candidate, we folded the raw data using \texttt{prepfold}, generating diagnostic plots that were inspected manually.

\subsection{Single-pulse Search}

To search for single pulses, the dedispersed time series was analyzed using the \texttt{single\_pulse\_search.py}.
The trial DM and DM steps used for the single-pulse search were the same as those for the periodicity search.
We identified single-pulse candidates with a signal-to-noise ratio (S/N) greater than 7 and folded the raw data according to the DM of each candidate using the \texttt{dspsr} software~\citep{2011PASA...28....1V}. 
Frequency-time plots were then generated for the candidates, which were visually inspected to confirm their validity.

\subsection{X-ray Observations}

\begin{figure*}
\begin{center}
\includegraphics[width=180mm]{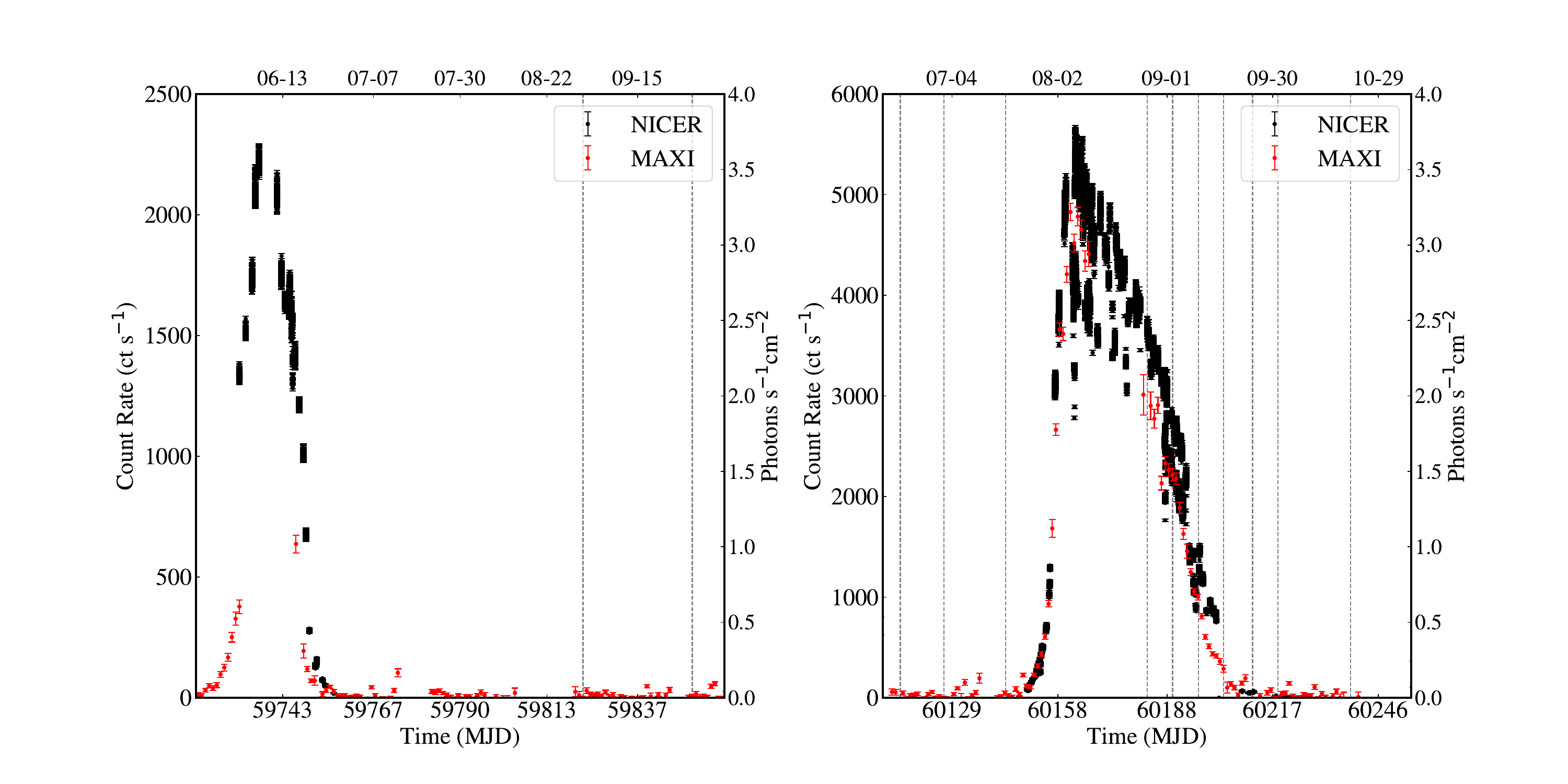}\\
\end{center}
\caption{The left and right panels display the X-ray light curves of Aql X-1 for the years 2022 and 2023, respectively. Black dots: NICER background-subtracted light curves in units of ct\,s$^{-1}$ (0.5--10.0\,keV, 16\,s resolution). Red dots: MAXI light curves in units of photons\,s$^{-1}$\,cm$^{-2}$ (2--20\,keV, 1 day resolution).
The gray dashed vertical lines indicate the MJDs of the FAST observations conducted in 2022 and 2023. }
\label{fig:obss}
\end{figure*}

We noted that Aql X-1 exhibited X-ray outbursts during 2022 and 2023, which were detected by the Neutron Star Interior Composition Explorer (NICER)~\citep{Alabarta2023ATel1, Alabarta2023ATel2, Alabarta2023ATel3, Homan2023}. 
To determine the precise time spans of the X-ray outburst and quiescence states, we retrieved the available NICER observational data for Aql X-1 from 2022 and 2023 through the High Energy Astrophysics Science Archive Research Center \footnote{\url{https://heasarc.gsfc.nasa.gov}}. 
The data were processed using HEASOFT 6.31.1 and NICER Data Analysis Software, applying standard filtering criteria for routine \texttt{NICERL2} to extract cleaned event data. 
The light curves of X-ray emission, binned at 16 s, were obtained using the \texttt{NICERL3-LC} pipeline.
We also retrieved the observational data of the Monitor of All-sky X-ray Image (MAXI), and obtained the 2–20 keV light-curve data with a bin size of 1 day from the MAXI Gas Slit Camera.\footnote{\url{http://maxi.riken.jp/mxondem/}}

\section{Results}
\label{sec:results}

\begin{figure}
    \centering
    \includegraphics[width=\columnwidth]{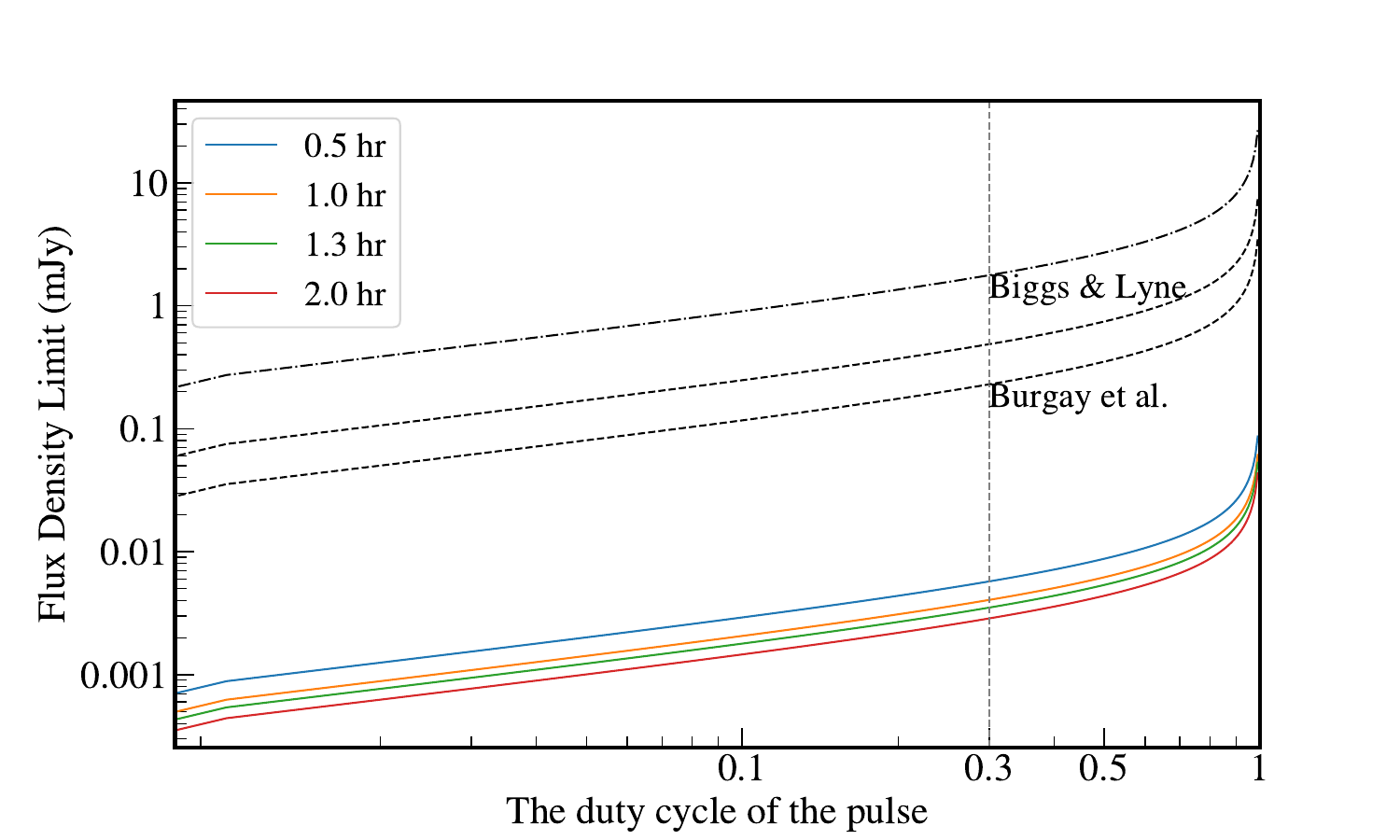}
    \caption{The upper limit flux density for Aql X-1 as a function of the duty cycle. Solid lines in different colors represent various integration times. The dotted-dashed and dotted lines indicate the limits obtained by \citet{Biggs1996} and \citet{Burgay2003}, respectively, with the flux densities scaled to 1250 MHz assuming a spectral index of $-1.6$. The gray dashed line denotes the duty cycle of the pulse as utilized in our work.}
    \label{fig:duty}
\end{figure}

In Figure~\ref{fig:obss}, we present the light curves for Aql X-1 observed by NICER \and MAXI between 2022 and 2023, with high and low backgrounds subtracted. 
The observations with FAST are marked as vertical dotted lines, covering both the X-ray burst and quiescence states. 
Using the ephemerides of Aql X-1 provided by~\citet{Mata2017}, we calculated the orbital phase of each observation by assuming a circular orbit (see the fifth column of Table~\ref{tab:observation}). 
Our observations span different orbital phases. 
We performed periodic and single-pulse searches for each observation, but no pulsed signals were detected. 

We estimated the upper limit of the flux density for each observation. 
For the periodicity search, the minimum detectable flux density is given by~\citep{Lorimer2004}
\begin{equation}
\label{eq1}
S_{\min} = \frac{(S/N)\beta T_{\rm sys}}{G(n_{\rm p}t_{\rm int}\Delta F)^{1/2}} \left(\frac{W}{P-W}\right)^{1/2},
\end{equation}
where \( G \) is the telescope gain, \( T_{\rm sys} \) is the system temperature, \( n_{\rm p} \) is the number of polarizations, \( \Delta F \) is the observational frequency bandwidth, \( \beta \) is the sensitivity degradation factor, \( W \) is the pulse width, \( P \) is the spin period, and \( S/N \) is the threshold signal-to-noise ratio required for detection. 
For FAST observations, the parameters are \( G = 16~\mathrm{K \,Jy^{-1}} \), \( T_{\rm sys} = 24~\mathrm{K} \), \( n_{\rm p} = 2 \), \( \beta = 1 \), and \( \Delta F = 400~\mathrm{MHz} \)~\citep{Jiang2020}. 

As described in Section~\ref{sec:obs}, the observed pulse width \( W_{\rm obs} \) depends on the intrinsic pulse width, interstellar scattering, and smearing across individual frequency channels~\citep{Cordes2003}. 
The effects of pulse smearing and the receiver's time resolution are negligible, but \( t_{\rm scatt} \) strongly depends on the DM. 
Figure~\ref{fig:duty} shows how the upper limit of the flux density varies with the pulse duty cycle, showing that the upper limit increases as the duty cycle increases. 
Assuming \( W = 0.3P \), which is typical for MSPs, and \( S/N = 7 \), we estimated \( S_{\min} \) for each observation. 
The results, shown in the last column of Table~\ref{tab:observation}, indicate that \( S_{\min} \) ranges from \( \sim 2.86 \) to \( 5.73~\mu\mathrm{Jy} \), depending on the duration of each observation. 

For the single-pulse search, the minimum detectable peak flux density is~\citep{Cordes2003}
\begin{equation}
\label{eq2}
S_{\min} = \frac{(S/N_{\rm peak}) 2 \beta T_{\rm sys}}{G(n_{\rm p} W \Delta F)^{1/2}},
\end{equation}
where \( S/N_{\rm peak} \) is the peak signal-to-noise ratio of a pulse. 
Single-pulses from a given pulsar are typically unstable, exhibiting varying widths~\citep[e.g.,][]{Parthasarathy2021, Wang2024}. 
Assuming a single-pulse width of \( W = 0.3P \) (about 0.54\,ms) and \( S/N_{\rm peak} = 7 \), we derived an upper limit for the detectable peak flux density of \( 15.6~\mathrm{mJy} \). 
It should be noted that this value has significant uncertainty.

\section{Discussion and Conclusions}
\label{sec:con}

\begin{table*}
\centering
\footnotesize
\caption{$L_{\rm X}$ for tMSPs in different states}.
\renewcommand{\arraystretch}{1.2} 
\setlength{\tabcolsep}{4pt}
\begin{tabular}{cccccccc}
\hline

Name &  \multicolumn{3}{c}{Disk State}    & Rotation-powered State  & References \\
 &   $L_{\rm X, flare}$  &  $L_{\rm X, high}$  &  $L_{\rm X, low}$  & $L_{\rm X, rot}$ \\
 &   ($10^{33} \,\rm  erg/s$)  &  ($10^{33} \, \rm erg/s$)  & ($10^{33} \, \rm erg/s$)  & ($10^{33} \, \rm  erg/s$) \\
 \hline
PSR J1023+0038 & 9.6(1)  & 2.85(4) & 0.57(3) & 0.16(5) & \citet{Archibald2015,2014ApJ...795...72L}\\
XSS J12270$-$4859 & 9.6(1) & 4.2(1) & 0.64(1) & 0.11(1) & \citet{Miraval2020}\\
IGR J18245$-$2452 & ...  &3.9(1) & 0.56(10) & 0.22(4) & \citet{Linares2014} \\
  &  & Propeller & Radio-ejection & Radio pulsar &  \citet{Campana2016aa,Miraval2020}  \\
\hline
\end{tabular}
\label{tab:tmsps}
\end{table*}

We present radio observations of Aql X-1 during both X-ray outburst and quiescence states. 
No pulsed signals were detected by FAST through periodicity and single-pulse searches. 
The lowest upper limit on the flux density from our observations is \( \sim 2.86~\mu\mathrm{Jy} \), assuming a pulse width of \( W = 0.3P \). 
Previous searches for radio emission from Aql X-1 were conducted using the 76-m Lovell Telescope at 925 MHz~\citep{Biggs1996} and the Parkes 64-m radio telescope at 1.4 GHz~\citep{Burgay2003}, and no pulsed signals were detected. 
Note that at that time, the rotation period of Aql X-1 was not yet determined. 
Pulsars typically exhibit steep spectra that can be modeled by a simple power law~\citep{Sieber1973}. 
Using a mean spectral index of \( -1.6 \)~\citep{Lorimer1995}, we estimated the implied upper limit flux densities for Aql X-1 using the same method from~\citet{Biggs1996} and~\citet{Burgay2003}, as shown in Figure~\ref{fig:duty}. 
Our results are more than ten times lower than previous limits.

For AMXPs, during X-ray outburst states, the magnetospheric radius is smaller than the corotation radius. 
In such cases, accretion matter reaches the neutron star surface, producing X-ray emission, and quenches the pulsar's radio emission, making it undetectable~\citep{Patruno2021}. 
During quiescence, the radio pulsar is expected to be active according to the recycling scenario~\citep{Patruno2021}, supported by optical observations~\citep{2003A&A...404L..43B}.
However, despite many searches, no radio pulsations have been detected in AMXPs~\citep{Burgay2003,Iacolina2009,Iacolina2010}.

For Aql X-1 in quiescence, it is unclear whether the system is in the radio-ejection phase or the radio-pulsar phase. 
In the radio-pulsar phase, nondetection could arise from several factors: (1) The radio emission may be intrinsically too weak. However, Aql X-1's distance of \( \sim 4.5 \) kpc is smaller than that of IGR J18245$-$2452 (\( \sim 5.5 \) kpc) and comparable to other known radio MSPs, making this explanation less likely. (2) The radio beam may not sweep across Earth. Radio beams are generally narrower than X-ray beams in MSPs (see \citealt{Lorimer2008} for a review). 
\citet{Burgay2003} estimated that the average sky fraction swept by radio beams for MSPs is \( \sim 0.57 \).  Given that radio pulsations are detected in only one AMXP (IGR J18245$-$2452) among over 20 known AMXPs~\citep{Patruno2021}, this scenario is unlikely to be the primary cause of nondetection in a statistical sense. However, for Aql X-1, it is unknown whether the radio beam is pointed away from Earth. (3) The radio pulsed emission may be intermittent. In this scenario, the radio pulsed emission may occur at particular times or orbital phases. 
IGR J18245$-$2452 was sporadically detected as a faint radio pulsar with a flux density of $10-20\,{\rm \mu Jy}$ at 2 GHz~\citep{2006MsT..........5B,Papitto2013}. 
We conducted only 12 observations of Aql X-1, with a cadence ranging from several days to tens of days. This is insufficient to provide constraints on this scenario. Further observations with smaller cadences and longer durations are needed.

In the radio-ejection phase, the radio pulsar remains active but is undetectable due to absorption by surrounding material in the radio-ejection state~\citep{Burderi2001}. 
The discovery of tMSPs has provided further insight into the behavior of AMXPs in quiescence~\citep{Archibald2009,Papitto2013,Bassa2014}. TMSPs exhibit different X-ray modes during the disk state~\citep{2018ASSL..457..149C}, as shown in Table~\ref{tab:tmsps}.
To explain the various modes of tMSPs during the disk state, \citet{Campana2016aa} proposed the propeller model, in which the inflowing disk matter is halted at the magnetospheric boundary, acting as a centrifugal barrier. According to this model, a small fraction of disk matter may leak through the centrifugal barrier~\citep{Campana2016aa} and reach the neutron star surface due to an advection-dominated accretion flow, where radiation is trapped within the disk~\citep{1994ApJ...428L..13N}. This process corresponds to the high mode of X-ray luminosity~\citep{Campana2016aa,Miraval2020}. If the accretion rate decreases, the disk pressure diminishes, causing the magnetospheric radius to expand. When the magnetospheric radius reaches the light cylinder radius, the system transitions into the radio-ejection phase, where accreted matter is expelled, corresponding to the low mode of X-ray luminosity~\citep{Burderi2001,Campana2016aa}. 
If the mass accretion rate decreases further, the system enters the radio pulsar state, where the corresponding X-ray luminosity is even lower~\citep{Campana2016aa}. 
For Aql X-1, the quiescent X-ray luminosity, $L_{\rm X} \sim 0.6 \times 10^{33}\,{\rm erg\,s^{-1}}$~\citep{1998ApJ...499L..65C}, is comparable to that of tMSPs in the low mode. 
By analyzing the X-ray luminosity and hard spectrum of Aql X-1 in quiescence, \citet{1998ApJ...499L..65C} suggested that its observational properties can be interpreted as shock emission from a reactivated rotation-powered pulsar. 
These phenomena collectively support the hypothesis that Aql X-1 is in the radio-ejection phase during quiescence. 
The radio pulsed emission of Aql X-1 could potentially be detected if the mass accretion rate decreases further in quiescence.

\citet{Burgay2003} investigated the surrounding matter under the assumption that Aql X-1 is in the radio-ejection phase during quiescence and found that radio pulses at 1.4 GHz could be easily absorbed by this material. 
However, the precise frequency at which the radio pulsed emission might become detectable remains uncertain due to the limited understanding of the surrounding matter’s properties. 
Future observations of AMXPs at higher frequencies could provide valuable insights into their nature, utilizing facilities such as the QiTai radio Telescope~\citep{Wang2023}, Atacama Large Millimeter/submillimeter Array~\citep{2018PASP..130a5002M}, Next Generation Very Large  Array, and Square Kilometre Array~\citep{selina2018,braun2019}. 
If radio pulsed emissions are detected at high frequencies but not at low frequencies, this would serve as direct evidence supporting the radio-ejection model.
Additionally, observations using radio arrays could help constrain the radio continuum flux, which is essential for advancing our understanding of the accretion processes in AMXPs.

\section*{Acknowledgments}
We thank the referee for the valuable comments, which improved our manuscript. We acknowledge the science research grants from the China Manned Space Project. This work is supported by the National Natural Science Foundation of China (Nos. 12288102, 12103042, 12203092, 12041304, 12273030, and 12473041), the science and technology innovation Program of Hunan Province (No. 2024JC0001), the Major Science and Technology Program of Xinjiang Uygur Autonomous Region (No. 2022A03013-3), the National SKA Program of China (No. 2020SKA0120100), the National Key Research and Development Program of China (Nos. 2022YFC2205202 and 2021YFC2203502), the Natural Science Foundation of Xinjiang Uygur Autonomous Region (No. 2022D01B71), and the Tianshan Talent Training Program for Young Elite Scientists (No. 2023TSYCQNTJ0024). This work made use of the data from FAST (Five-hundred-meter Aperture Spherical radio Telescope; https://cstr.cn/31116.02.FAST). FAST is a Chinese national megascience facility, operated by National Astronomical Observatories, Chinese Academy of Sciences. The research is partly supported by the Operation, Maintenance and Upgrading Fund for Astronomical Telescopes and Facility Instruments, budgeted from the Ministry of Finance of China (MOF) and administrated by the Chinese Academy of Sciences (CAS).

 \software{PRESTO \citep{Ransom2001}, DSPSR \citep{2011PASA...28....1V}, HEASOFT(v6.31.1) , NICERDAS(v010a)}

\bibliography{sample01}{}
\bibliographystyle{aasjournal}

\end{document}